\documentclass[12pt,twoside,a4paper]{article}
\usepackage{amsmath} \usepackage{amsfonts} \usepackage{amssymb}
\usepackage{romp}

\def\be{\begin{equation}}
\def\ee{\end{equation}}
\def\ba{\begin{array}}
\def\ea{\end{array}}

\def\Cb{\ \hbox{\vrule width 0.6pt height 6pt depth 0pt
              \hskip -3.2 pt} C}
\def\qed{\rule{5pt}{5pt}}

\title{Addendum to ``Multipartite states under local unitary transformations''}
\author{ Sergio Albeverio\thanks{ SFB 611; IZKS; BiBoS;
CERFIM (Locarno); Acc. Arch. USI (Mendrisio)} \\ Institut f\"ur
Angewandte Mathematik,
Universit\"at Bonn, Wegelerstr. 6, D-53115 Bonn \\ e-mail: albeverio@uni-bonn.de  \\[2ex]
         Laura Cattaneo \\ Institut f\"ur
Angewandte Mathematik, Universit\"at Bonn, Wegelerstr. 6, D-53115
Bonn \\ e-mail: cattaneo@wiener.iam.uni-bonn.de \\[2ex]
         Shao-Ming Fei \\ Institut f\"ur
Angewandte Mathematik, Universit\"at Bonn, Wegelerstr. 6, D-53115
Bonn \\ Department of Mathematics, Capital Normal University,
Beijing 100037 \\ Max Planck Institute for Mathematics in the
Sciences, Inselstr. 22, D-04103 Leipzig \\ e-mail: fei@uni-bonn.de \\[2ex]
         Xiao-Hong Wang \\ Department of Mathematics, Capital Normal University,
Beijing 100037 \\ e-mail: wangxh@mail.cnu.edu.cn}

\begin{document}

\maketitle

\begin{abstract}
In previous work the authors introduced a notion of generic states
and obtained criteria for local equivalence of them. Here they
introduce the concept of CHG states maintaining the criteria of
local equivalence. This fact allows the authors to halve the number
of invariants necessary to characterize the equivalence classes
under local unitary transformations for the set of tripartite states
whose partial trace with respect to one of the subsystems belongs to
the class of CHG mixed states.
\end{abstract}

\noindent {\bf Keywords:}  tripartite quantum states, local unitary
transformations, entanglement, invariants\\

In the paper \cite{acfw2} we exploited the equivalence criterion
for the members of a class of bipartite mixed states constructed
in \cite{afw,afpy} to write an equivalence criterion for the
members of a class of pure tripartite states. In this Addendum we
show how
the conditions assumed in \cite{afw,afpy,acfw2} can be relaxed.\\

We first consider (mixed) states on a bipartite system $H_A\otimes
H_B$, where $H_A$ and $H_B$ are finite dimensional Hilbert spaces
of dimension $N_A$ and $N_B$, respectively. Let $\rho$ be a
density matrix defined on $H_A\otimes H_B$ with $rank(\rho)=n\leq
N^2$, where $N=\min\{N_A,N_B\}$. $\rho$ can be decomposed
according to its eigenvalues and eigenvectors:
$$
\rho=\sum_{i=1}^n\lambda_i\vert\varphi_i\rangle\langle\varphi_i\vert,
$$
where $\lambda_i$ resp. $\vert\varphi_i\rangle$, $i=1,...,n$, are
the nonzero eigenvalues resp. eigenvectors of the density matrix
$\rho$. $\vert\varphi_i\rangle$ has the form
\[\vert\varphi_i\rangle=\sum_{k=1}^{N_A} \sum_{l=1}^{N_B} a_{kl}^i
\vert e_k\rangle\otimes \vert f_l\rangle,~~~ a_{kl}^i\in\Cb,~~~
\sum_{k=1}^{N_A}\sum_{l=1}^{N_B} a_{kl}^i a_{kl}^{i\ast}=1,~~~
i=1,...,n\,,\] where $\{|e_i\rangle\}_{i=1}^{N_A}$ and
$\{|f_i\rangle\}_{i=1}^{N_B}$ are orthonormal bases in
$\mathcal{H}_A$ and $\mathcal{H}_B$, respectively, and $^*$ means
complex conjugation. Let $A_i$ denote the matrix given by
$(A_i)_{kl}=a_{kl}^i$. We introduce $ \left\{\rho_{i}\right\}$,  $
\left\{\theta_{i}\right\}$:
\begin{equation} \rho_i=Tr_B
\vert\varphi_i\rangle\langle\varphi_i\vert=A_iA_i^\dag,~~~
\theta_i=(Tr_A
\vert\varphi_i\rangle\langle\varphi_i\vert)^\ast=A_i^\dag A_i, ~~~
i=1,...,n, \label{rti}
\end{equation}
with $^\dag$ denoting adjoint. $Tr_A$ and $Tr_B$ stand for the
traces over the first and second Hilbert spaces respectively, and
therefore, $\rho_i$ and $\theta_i$ can be regarded as reduced
density matrices. Let $\Omega(\rho)$ and $\Theta(\rho)$ be two
``metric tensor" matrices, with entries given by
\begin{equation}
\Omega(\rho)_{ij}=Tr(\rho_i\rho_j),~~~
\Theta(\rho)_{ij}=Tr(\theta_i\theta_j),~~~{\rm for}~ i,j=1,...,n,
\label{ij}
\end{equation}
and
$$
\Omega(\rho)_{ij}=\Theta(\rho)_{ij}=0,~~~{\rm for}~ N^2\geq i,j>n.
$$
In \cite{afpy} a mixed state $\rho$ is called {\it generic} if the
corresponding ``metric tensor" matrices $\Omega$ and $\Theta$
satisfy
\begin{equation}
\det(\Omega(\rho))\neq
0\quad\mathrm{and}\quad\det(\Theta(\rho))\neq 0. \label{gen}
\end{equation}
We shall say here that a mixed state $\rho$ is {\it high generic}
if
\begin{equation}
\det(\Omega(\rho))\neq
0\quad\mathit{or}\quad\det(\Theta(\rho))\neq 0. \label{wgen}
\end{equation}
If we add conditions: \be\label{comm} [\rho_i, \rho_j]=0,
~~[\theta_i, \theta_j]=0\ee and $\rho_i$ is a full rank matrix. We
call a mixed state is a {\it comm-high generic} or {\it CHG} state
if it is a high generic one and also satisfy the above two
conditions. Condition (\ref{comm}) assures that $A_iA_i^\dag$ and
$A_i^\dag A_i$ have common eigenvectors.

Similarly we also introduce trilinear expressions $X(\rho)$ and
$Y(\rho)$ as
\begin{equation}
X(\rho)_{ijk}=Tr(\rho_i\rho_j\rho_k),~~~
Y(\rho)_{ijk}=Tr(\theta_i\theta_j\theta_k),~~~~ i,j,k=1,...,n.
\label{ijk}
\end{equation}


\begin{theorem}{Proposition} Two CHG density matrices with non-degenerate
$\Omega$ (or non-degenerate $\Theta$) are equivalent under
local unitary transformations if and only if there exists an
ordering of the corresponding eigenstates such that the following
invariants have the same values for both density matrices:
\begin{eqnarray}
&& J^s(\rho) = Tr_B(Tr_A\rho^s),\quad s=1,...,n\nonumber\\
&&\Omega(\rho),\; X(\rho)\quad (or\;\Theta(\rho),\; Y(\rho))
\label{theorem}
\end{eqnarray}
\label{theo1}
\end{theorem}

\noindent{\it Proof:} In \cite{afw} it was proved that two generic
states such that $\Omega$ and $\Theta$ are both non-degenerate are
equivalent under local unitary transformations if and only if there
exists an ordering of the corresponding eigenstates such that the
invariants $J^s(\rho)$, $s=1,\dots,n$, $\Omega(\rho)$,
$\Theta(\rho)$, $X(\rho)$, and $Y(\rho)$ take the same values for
both density matrices. In particular (see \cite[eq.(14)]{afw} or
\cite[eq.(15)]{afpy}), from the conditions $\Omega(\rho)$
non-degenerate, $\Omega(\rho)=\Omega(\rho')$, and $X(\rho)=X(\rho')$
follows that
\begin{equation}
\rho_i^\prime=u\rho_i u^\dag, \label{f2}
\end{equation}
for some $u\in\mathcal{U}_A$, where $\mathcal{U}_A$ denotes the
space of all unitary matrices on $\mathcal{H}_A$. So we have
$A_{i}^\prime A_{i}^{\prime\dag}= u A_{i} A_{i}^{\dag} u^\dag$. Thus
$A_{i} $ and $A_{i}^\prime$ have the same singular values. The {\it
Singular value decomposition} of matrices (see, e.g.,\cite{svd}) and
(\ref{comm}) assure the existence of unitary matrices $U,V,U^\prime,
V^\prime$ such that
$$UA_{i}V=diag(s_1(A_{i}),\cdots, s_n(A_{i})),~~~
U^\prime A_{i}^\prime V^\prime=diag(s_1(A_{i}^\prime),\cdots,
s_n(A_{i}^\prime)),$$ where $s_j(A_{i})$ and $s_j(A_{i}^{\prime})$
represent the $j$-$th$ singular value of $A_{i}$ and $A_{i}^\prime$,
respectively, and $diag$ means the principal diagonal (of a non
necessarily diagonal matrix). Because of
$s_j(A_{i}^\prime)=s_j(A_{i})$ for all $i$, we have $ A_{i}^\prime
=u_1 A_{i} w_1$, and $\vert\varphi_i^\prime\rangle=u_1\otimes w_1
\vert\varphi_i\rangle$, $i=1,...,n$, where $u_1={U^\prime}^\dag U,
w_1={V^\prime}^*V^T.$

From the genericity condition $\Theta(\rho)$ non-degenerate, and
from $\Theta(\rho)=\Theta(\rho^{\prime})$ and
$Y_{ijk}(\rho)=Y_{ijk}(\rho^{\prime})$ we can similarly deduce
that
\begin{equation}
\theta_i^\prime=w^{\dag}\theta_i w \label{f3}
\end{equation}
for some $w\in\mathcal{U}_B$. Again from the {\it Singular value
decomposition} of matrices, we have
$\vert\varphi_i^\prime\rangle=u\otimes w \vert\varphi_i\rangle$,
$i=1,...,N^2$, and $\rho^\prime=u\otimes w ~\rho~u^\dag\otimes
w^\dag$. Hence $\rho^\prime$ and $\rho$ are equivalent under local
unitary transformations: the common $u$ and $v$ for different
$\vert\varphi_i\rangle$'s can be obtained in two ways, either from
the condition $\Omega(\rho)$ non-degenerate,
$\Omega(\rho)=\Omega(\rho^{\prime})$ and
$X_{ijk}(\rho)=X_{ijk}(\rho^{\prime})$, or from $\Theta(\rho)$
non-degenerate, $\Theta(\rho)=\Theta(\rho^{\prime})$ and
$Y_{ijk}(\rho)=Y_{ijk}(\rho^{\prime})$. Therefore, it is
sufficient to consider only $\Omega(\rho)$ and $X(\rho)$ (or
$\Theta(\rho)$ and $Y(\rho)$).  \qed\\

Consider now tripartite pure states of a system $H_A\otimes
H_B\otimes H_C$, where $H_i$, $i\in\{A,B,C\}$, is a finite
dimensional Hilbert space of dimension $N_i$. By using the results
of Proposition \ref{theo1}, Proposition 2 in \cite{acfw2} can be
rewritten as follows.

\begin{theorem}{Proposition}
Let $|\psi\rangle$ and $|\psi'\rangle$ be two pure states of
$\mathcal{H}_A\otimes\mathcal{H}_B\otimes\mathcal{H}_C$ and assume
that $\rho = \operatorname{Tr}_A(|\psi\rangle\langle\psi|)$ is a CHG
mixed state. $|\psi\rangle$ is equivalent to $|\psi'\rangle$ under
local unitary transformations if and only if
\begin{equation}
I_{\alpha,\beta}^{A,s}(|\psi\rangle)=I_{\alpha,\beta}^{A,s}(|\psi'\rangle)
\label{Inv3}
\end{equation}
for $s\in\{B,C\}$,
$\alpha=1,\dots,\operatorname{min}\{N_B^2,N_C^2\}$,
$\beta=1,\dots,N_r$, where $r\in\{B,C\}$ but is different from
$s$, and for $\rho' =
\operatorname{Tr}_A(|\psi'\rangle\langle\psi'|)$
\begin{eqnarray*}
&& \Omega(\rho)_{jk} =\Omega(\rho')_{jk}\,,\quad
X(\rho)_{jkl}=X(\rho')_{jkl}\,,\\
&& (or\;\Theta(\rho)_{jk}=\Theta(\rho')_{jk}\,,\quad
Y(\rho)_{jkl}=Y(\rho')_{jkl})\,,
\end{eqnarray*}
for the $j,k$ such that $\lambda_j=\lambda_k$. \label{prop2}
\end{theorem}

The weaker conditions allow us to halve the number of invariants
to consider, as compared to the statements of \cite{acfw2}.\\

\vspace{0.5cm} \noindent {\bf Acknowledgments}\\
L. Cattaneo gratefully acknowledges the financial support by the
Lise Meitner Fellowship of the Land Nordrhein-Westfalen. X.-H. Wang
gratefully acknowledges the support provided by the China-Germany
Cooperation Project 446 CHV 113/231, ``Quantum information and
related
mathematical problems".\\


\vspace{0cm}

\begin{thebibliography}{99}
\bibitem{acfw2} S. Albeverio, L. Cattaneo, S.M. Fei, and X.H. Wang: Multipartite states under local unitary transformations,
{\it Rep. Math. Phys.} {\bf 56} (2005), 341.
\bibitem{afw} S. Albeverio, S.M. Fei, and X.H. Wang: Equivalence of bipartite quantum mixed states under local unitary
transformations, in {\em Proc. First Sino-German Meeting on
Stochastic Analysis - Satellite Conference to the ICM 2002\/}, Edts.
S. Albeverio, Z.M. Ma, and M. R\"ockner, Beijing 2002.
\bibitem{afpy} S. Albeverio, S.M. Fei, P. Parashar,
and W.-L. Yang: Nonlocal properties and local invariants for
bipartite systems, {\it Phys. Rev. A\/} {\bf 68} (2003), 010313.
\bibitem{svd} G.H. Golub and C.F. Van Loan: {\it Matrix Computations}, 3rd ed., Johns Hopkins University Press, Baltimore
1996.


\end{thebibliography}
\end{document}